# Direct imaging of a zero-field target skyrmion and its polarity switch in a chiral magnetic nanodisk

*Fengshan Zheng[1], Hang Li[2], Shasha Wang[3,4], Dongsheng Song[5], Chiming Jin[3,4], Wenshen Wei[3,4], András Kovács[1], Jiadong Zang[2]\*, Mingliang Tian[3,4], Yuheng Zhang[3,4], Haifeng Du[3,4]\*, and Rafal E. Dunin-Borkowski[1]*

[1] Ernst Ruska-Centre for Microscopy and Spectroscopy with Electrons and Peter Grünberg Institute, Forschungszentrum Jülich, 52425 Jülich, Germany

[2] Department of Physics, University of New Hampshire, Durham, New Hampshire 03824, USA

[3] The Anhui Key Laboratory of Condensed Matter Physics at Extreme Conditions, High Magnetic Field Laboratory, Chinese Academy of Sciences and University of Science and Technology of China, Hefei 230026, China

[4] Collaborative Innovation Center of Advanced Microstructures, Nanjing University, Jiangsu Province 210093, China

[5] National Center for Electron Microscopy in Beijing, Key Laboratory of Advanced Materials (MOE) and the State Key Laboratory of New Ceramics and Fine Processing, School of Materials Science and Engineering, Tsinghua University, Beijing 100084, China

**\*Corresponding authors**: *Jiadong.Zang@unh.edu*; *duhf@hmfl.ac.cn.*

A target skyrmion is a flux-closed spin texture that has two-fold degeneracy and is promising as a binary state in next generation universal memories. Although its formation in nanopatterned chiral magnets has been predicted, its observation has remained challenging. Here, we use off-axis electron holography to record images of target skyrmions in a 160-nm-diameter nanodisk of the chiral magnet FeGe. We compare experimental measurements with numerical simulations, demonstrate switching between two stable degenerate target skyrmion ground states that have opposite polarities and rotation senses and discuss the observed switching mechanism.

**Introduction**

Recent studies of nanoscale topological spin textures have changed the landscape of magnetism [1]. Magnetic skyrmions, which are characterized by topological indices, exhibit novel dynamics and are of interest as carriers of binary digits or logic elements in future spintronic devices [2–5]. The engineering difficulty and energy consumption of such devices would be reduced by the realization of zero-field skyrmion states [6,7]. Unfortunately, the stabilization of skyrmions in chiral magnets usually results from a combination of ferromagnetic exchange, anti-symmetric chiral Dzyaloshinskii-Moriya (DM) interactions and the presence of an externally applied magnetic field, with the first two coupling terms giving rise to a spin helix ground state [8]. However, a zero-field skyrmion is predicted to be possible in a nanopatterned structure because the magnetostatic energy then prefers a flux-closed state such as a magnetic skyrmion [9]. Such a spin configuration would offer the prospect of a purely electrical skyrmion-based device. As a result, there is great interest in the study of skyrmion physics in confined geometries [4,10,11]. In particular, theoretical work on nanodisks has suggested the possibility of forming an exotic topological texture termed a target skyrmion [10–16]. However, the target skyrmion has not yet been observed experimentally.

A major obstacle that hinders the imaging of the magnetic texture of a target skyrmion is the fact that the size of the nanodisk that supports it must be comparable to the size $a_{sk}$ of the skyrmion itself, which is usually on the deep-submicron scale [10–14]. In a previous report [17], Lorentz transmission electron microscopy

(TEM) was used to study chiral magnetic FeGe nanodisks with diameters $d > 180\ nm$, that were approximately twice as large as $a_{sk} \sim 80$ nm for FeGe. The ground state in these nanodisks was a spin helix [17]. The study of smaller samples has been hindered both by sample preparation and by limitations in the spatial resolution of magnetic imaging techniques. For example, it is difficult to measure spin textures in nanostructures using Lorentz TEM because of the presence of Fresnel fringes, which are affected by local changes in sample thickness and composition [18,19]. Such unwanted contributions to the recorded contrast can be eliminated more easily when using the TEM technique of off-axis electron holography (EH), which has been used to provide direct access to electron-optical phase images of FeGe nanostructures with sizes of below 0.1 μm with nm spatial resolution [20,21]. Here, we use EH to study spin configurations in an FeGe nanodisk of diameter $d \sim 160$ nm. We identify zero-field target skyrmions and use a perpendicular applied magnetic field to reverse their polarities.

**Magnetic configuration of a target skyrmion**

A target skyrmion consists of a central skyrmion surrounded by one or more concentric helical stripes (Fig. 1) and can be regarded as a curved spin helical state [10, 22]. From the center of the disk to its boundary, the out-of-plane component of the magnetic moment rotates by an angle $\phi$ that is larger than the value of $\pi$ for a typical skyrmion. The parameter $\phi/\pi$ can then be used to characterize the number of half helical periods along any radial direction. As a result of the boundary confinement [10, 12–15], the outermost spin helix, termed as an edge twist, may correspond to an irrational fraction of a period. The value of $\phi$ is therefore not necessarily an integer multiple of $\pi$. The topological charge of a target skyrmion is also not integer-valued [23], and is given by the expression

$$Q = p(1 - \cos\phi)/2 \qquad (1)$$

Following definitions that are widely used to describe magnetic vortices in soft magnetic disks [24], we make use of two parameters - the polarity $p$ and the circularity $c$ - to describe the direction of the out-of-plane magnetization at the center of a target skyrmion and the rotational sense of its in-plane magnetization, respectively. The out-of-plane magnetization at the core can point either up ($p = +1$) or down ($p = -1$), while the in-plane magnetization can rotate clockwise ($c = +1$) or counterclockwise ($c = -1$). The circularity $c$ of the center skyrmion and the edge twist have opposite values [10, 11]. As a result of the fixed handedness and sign of the DM interaction, $c$ is determined uniquely once $p$ has been defined. In zero magnetic field, two degenerate configurations with opposite magnetization are possible (Figs 1a and 1b), as a result of the quadratic nature of the Heisenberg and DM interactions.

Here, we use EH to image the spin textures of individual target skyrmions. The technique is sensitive to the in-plane component of the magnetic induction integrated in the electron beam direction [21]. Simulations of the in-plane magnetization in ideal target skyrmions are shown in Figs 1c and 1d. In these images, the moments are normal to the plane of the disk in the skyrmion core and in the gap between the central skyrmion and the edge twist.

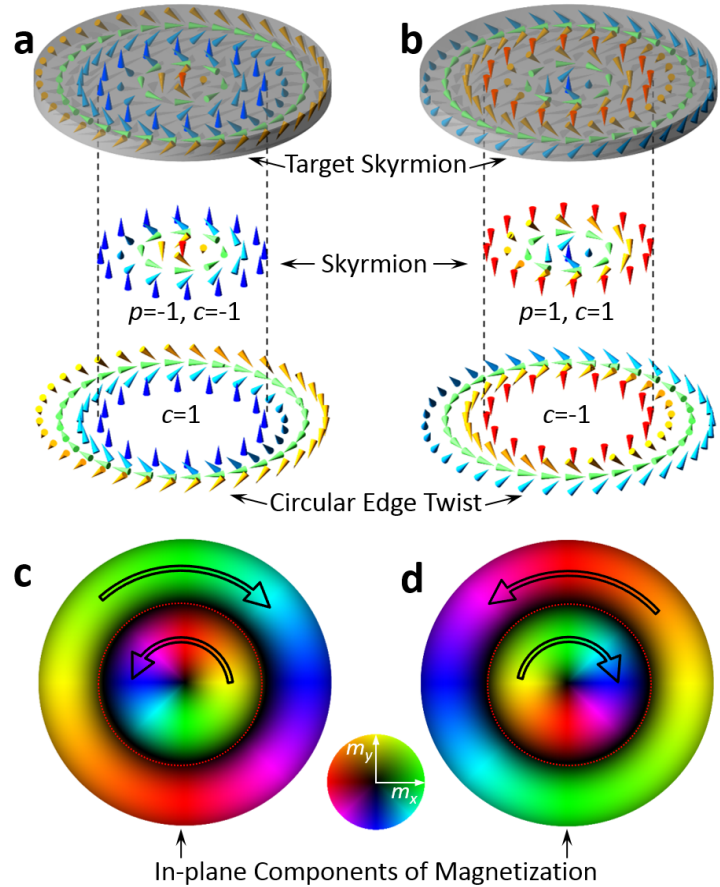

**Fig. 1: a** and **b**, Magnetization configurations of two-fold-degenerate target skyrmions for $p = -1$ and $p = +1$, respectively. **c** and **d**, In-plane magnetization in the target skyrmions in **a** and **b**, respectively with colors and arrows used to indicate the rotational sense of the in-plane magnetization, according to the color wheel shown. The black dashed lines in **a** and **b** and the thin red lines in **c** and **d** mark the boundaries between the skyrmions and the circular edge twists.

**Fabrication and magnetic imaging of a nanodisk**

Fabrication of a cylindrical nanodisk from a bulk crystal of FeGe was achieved using a lift-out method based on FIB milling (Figs 2a and 2b, and Supplementary Video 1). The nanodisk is embedded in amorphous $PtC_x$ and has a diameter of $d \sim 160$ nm and a thickness of $t \sim 90$ nm (Figs 2a and 2b). EH measurements of the projected in-plane magnetic induction were performed at a temperature of $T \sim 95$ K (see the Supplementary Material and Figs S1 and S2).

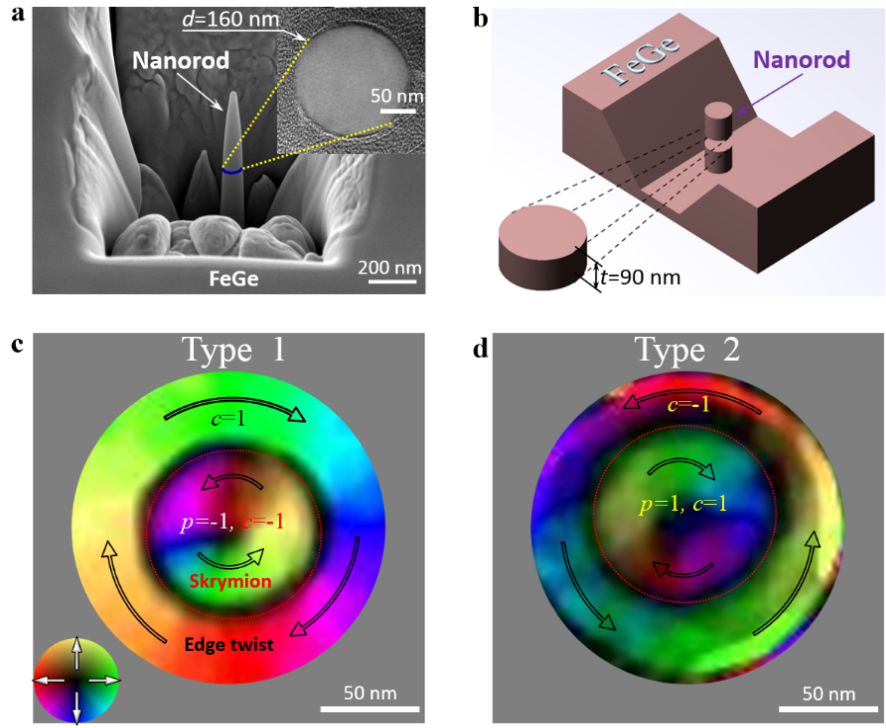

**Fig. 2: a** and **b**, SEM image and schematic illustration of the fabrication of an FeGe nanodisk using FIB milling. $t$ and $d$ are the thickness and diameter of the nanodisk, respectively. The inset in **a** shows a TEM bright-field image. **c** and **d**, Magnetic induction maps of spin texture in the nanodisk for opposite polarities of the spin state. Colors and arrows are used to indicate both the strength and the direction of the projected in-plane magnetic induction. The red lines show the boundary between the skyrmion and the edge twist.

Fig. 2c shows a representative magnetic induction map in the nanodisk recorded using EH after zero-field cooling (ZFC), which avoids possible metastable magnetic states [25]. When compared with the calculated in-plane magnetization shown in Fig. 1c, this image provides unequivocal evidence for a target skyrmion ground state.

The central region with circularity $c = -1$ is a complete skyrmion and is surrounded by an edge twist for which $c = +1$. The size of the central skyrmion is ~ 85 nm, which is close to the skyrmion lattice constant in thin film FeGe [26]. The polarity of the skyrmion cannot be measured directly, but can be inferred from its field-driven evolution, as discussed below. We label this target skyrmion Type 1. As mentioned above, target skyrmions with opposite polarity have the same energy in zero magnetic field. The same ZFC annealing procedure was used to obtain a magnetic state in which the central skyrmion corresponded to $c = +1$ (Fig. 2d). This target skyrmion is labeled Type 2. The size of the central skyrmion in the Type 2 is almost identical to that in the Type 1 configuration. However, the position of the central skyrmion deviates slightly from the center of the nanodisk. These differences are thought to result from pinning effects due to sample imperfections.

**Field-driven polarity reversal of a target skyrmion**

In order to switch between two degenerate target skyrmions, we applied a magnetic field **H** perpendicular to the disk. We define positive $H$ as pointing upwards (marked with the symbol ⊙ in Fig. 3). For the Type 1 target skyrmion (Fig. 2c), an increase in the magnetic field shrinks the central skyrmion, while both $c$ and $p$ are unchanged during the magnetization process until a saturated magnetic state is achieved at large $H$ (Fig. 3a). This observation is in agreement with the conventional field-driven evolution of a skyrmion in a two-dimensional thin film [3,8,26], in which a skyrmion is stable when an external magnetic field is applied antiparallel to the magnetization direction of its core. We conclude that the polarity of this target skyrmion is $p = -1$, as shown in Fig. 2c. Interestingly, the size of the central skyrmion exhibits a non-monotonic dependence on applied magnetic field and has a maximum value of 110 nm at $H \sim 134$ mT, which is almost 1.5 times $a_{sk} \sim 81$ nm (Fig. S4). This behavior is associated in part with the flexibility of the edge twist, which results from its non-quantized topological charge [5, 20].

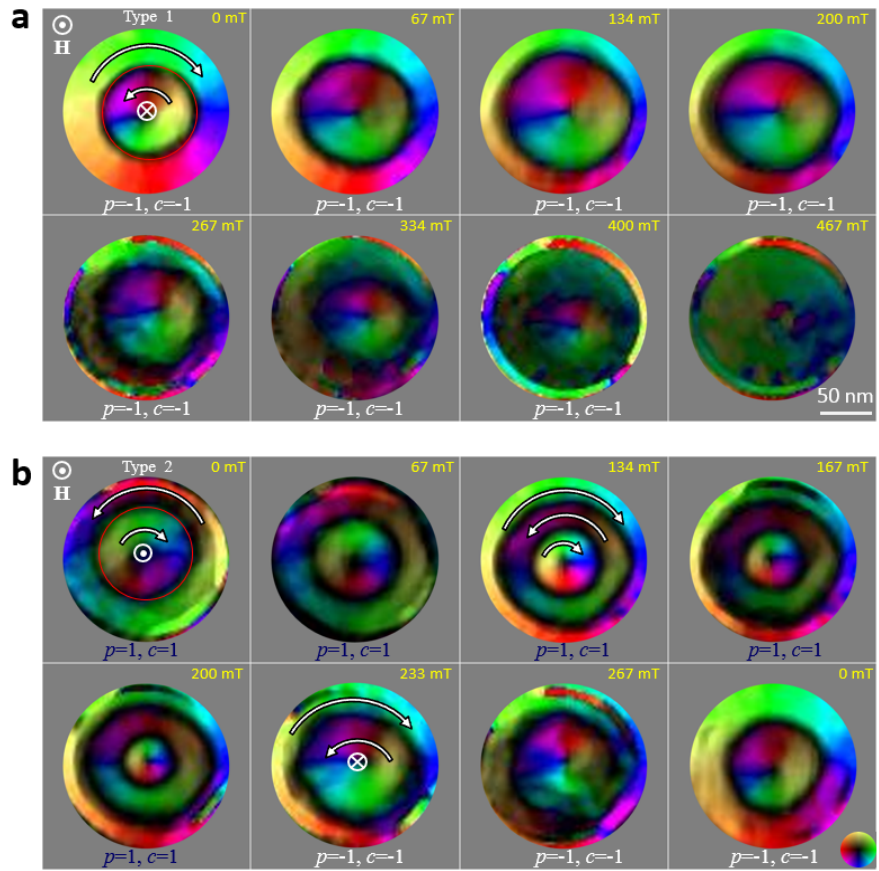

**Fig. 3:** Magnetic induction maps recorded using off-axis electron holography showing target skyrmions in an FeGe nanodisk for different applied magnetic fields. **a,** Target skyrmion of Type 1 with $p = -1$. The circularity and polarity persist during the entire magnetization process. **b,** Target skyrmion of Type 2 with $p = +1$. In an intermediate applied field, a new half-period curved spin helix forms between the central skyrmion and the edge twist (panel **b**, $H \sim 134$, 167 and 200 mT).

In a higher applied field, the target skyrmion evolves to the other type (panel **b**, $H \sim 233$ and 267 mT). The thick white arrows show the direction of rotation of the projected in-plane magnetic induction. ⊙ and ⊗ represent up and down orientations, respectively.

A different process is observed for a Type 2 target skyrmion (Fig. 3b). At low values of $H$ between 0 and 67 mT, the size of the central skyrmion decreases, while that of the edge twist increases. Above a threshold value of $H_{2\pi}$, the expansion of the edge twist cannot be sustained and it splits into a half-period spin helix and a new edge twist outside it (Fig. 3b; 134 mT). Not considering the new edge twist, the newly-formed spin texture is referred to as a *2π vortex* because the rotation angle $\phi$ from the center to the edge is exactly $2\pi$ [22]. The circularity ($c = -1$) in the new half-period helix is opposite to that of the central skyrmion, while for the new edge twist $c = +1$ (Fig. 3b; 134 mT). When the external field exceeds 200 mT, the central skyrmion shrinks and ultimately disappears. At the same time, the half-period circular helix takes over and evolves into a new skyrmion by collapsing its inner ring to a single point (Fig. 3b; 200 mT). The circularity of the newly-formed central skyrmion is $c = -1$, which follows the initial circular edge twist at $H = 0$ mT. $c$ is then opposite to that of the initial target skyrmion. As a result of the relationship between the polarity and the circularity, the polarity is also switched. The switching field, $H_s$, from an initial *2π* vortex to the new *π* vortex is $\sim 220$ *m*T. Once the polarity reversal of the target skyrmion has been achieved, a decrease in the field does not change the polarity (final image in Fig. 3b).

**Numerical simulation of magnetic-field-driven polarity reversal**

Our experimental results demonstrate two-fold degenerate zero-field target skyrmions and their field-driven polarity reversal. However, they only measure the projected in-plane magnetic induction [21]. As the disk has a thickness of 90 nm, which is comparable to $a_{sk}$ ~ 80 nm, a three-dimensional (3D) magnetization state may be supported [27–29]. We performed numerical simulations using a spin model for a 3D isotropic chiral magnet, in order to understand the spin arrangement and magnetic phase transitions in more detail. Parameters for the real sample were used in the simulation. Following a standard approach [12, 13], we first determined the lowest energy magnetic state by comparing the energies of typical equilibrium states relaxed from different initial states. The results show that the target skyrmion always has a lower energy than any other magnetic state, indicating that the ground state is consistent with the experimental results (Fig. S5).

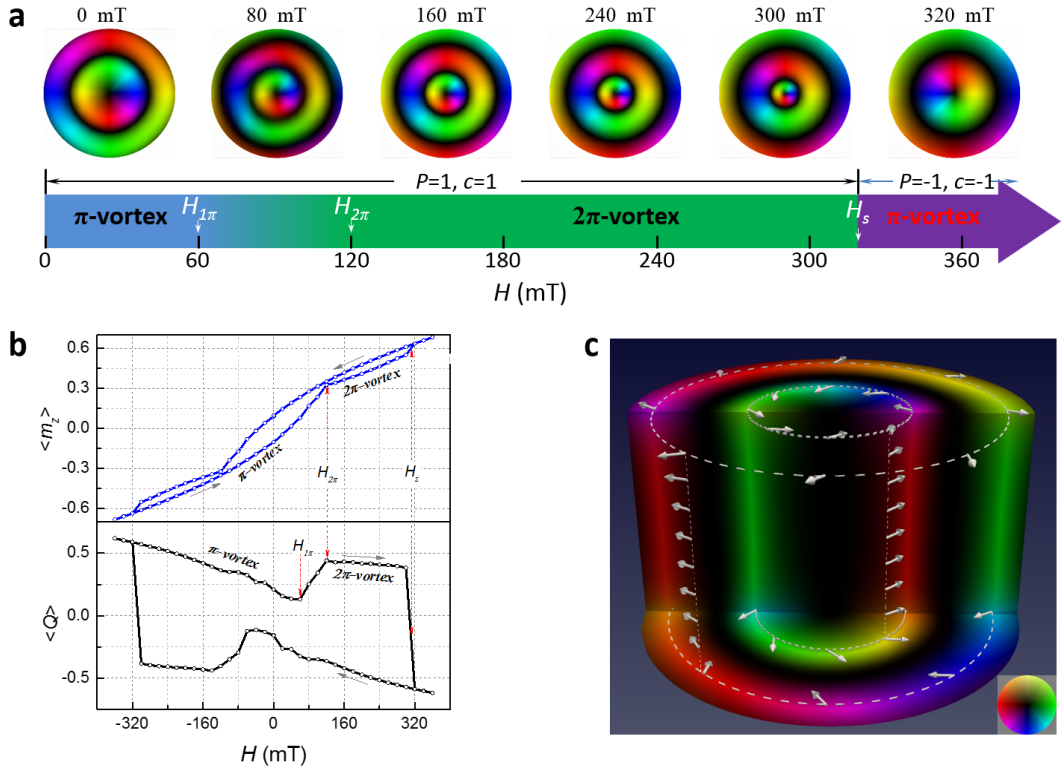

**Fig. 4:** Calculated in-plane magnetization in different applied magnetic fields. **a,** Magnetic field dependence of the in-plane magnetization of a Type 2 target skyrmion. **b,** Average magnetization $<m_z>$ and topological charge $<Q>$ plotted as a function of applied magnetic field. Three transition fields $H_{1\pi}$, $H_{2\pi}$ and $H_s$ are marked by red dotted lines. Gray arrows indicate the direction of the hysteresis loop. **c,** Three-dimensional spin arrangement of a target skyrmion in zero applied magnetic field.

As in the experiment results, the simulations followed the field-driven evolution of magnetic states starting from the two types of target skyrmion. For the Type 1 target skyrmion, the simulations show no switching during the entire magnetization process (Fig. S6). In contrast, a polarity switch is reproduced for the Type 2 target skyrmion. Fig 4a shows snapshots of the simulations during the switching process.

The applied magnetic field was extended further to the negative branch, allowing a complete hysteresis loop to be obtained, as shown in Fig. 4b. Just as in the experiment, a polarity reversal is identified at $H_s$. The transition field $H_s$ in the simulation (~ 320 mT) is larger than that in the experiment (~ 220 mT). This difference is attributed to the energy barrier for collapse of the central skyrmion, which can be overcome in real materials as a result of thermal fluctuations and defects. Collapse of the central skyrmion during this transition leads to dramatic changes in topology and magnetization, resulting in a decrease in topological charge and an increase in magnetization (Fig. 4b). As the reversal of the central magnetization during this transition switches the circularity, the sign of the topological charge changes, according to Eq. 1.

Below $H_s$, the transition from a $\pi$ vortex to a $2\pi$ vortex is observed, as reflected in the form of changes in real space spin state, magnetization and topological charge in Fig. 4b. In addition to a kink at $H_{2\pi}$ in both magnetization and topological charge, another kink $H_{1\pi}$ in the topological charge curve is observed. Climbing of <$Q$> between $H_{1\pi}$ and $H_{2\pi}$ suggests that the formation of a *$2\pi$ vortex* is not strictly a first-order phase transition. By examining the 3D magnetic structure in the transition region close to $H_{1\pi}$ (Fig. S7), we find that the expanded edge twist starts to detach from the boundary from the top to the bottom of the disk. This nucleation of a *$2\pi$ vortex* may explain the smooth change in magnetization close to $H_{1\pi}$. During this process the *z* component of the edge spin decreases, which explains the increase in $Q$ according to Eq. 1. In contrast, from $H = 0$ to $H_{1\pi}$, the expansion of the edge twist leads to an increase in the z component of the edge spin, $Q$ decreases and a kink is created at the transition field $H_{1\pi}$. This process is in full agreement with the experimental EH results.

The agreement between experimental results and theoretical simulations allows us to obtain insight into the mechanism of stability of the target skyrmion. Previous theoretical investigations have proposed that both the demagnetization energy [29, 30] and the magnetization variation normal to the disk plane [27–29] help to stabilize target skyrmions. These effects are confirmed by our analysis. The role of demagnetization energy is revealed in three ways. First, if the demagnetization energy is omitted, the target skyrmion cannot be achieved at equilibrium no matter what initial state is chosen (Fig. S5). Second, if we use the stabilized zero-field target skyrmion as the initial state and then turn off the demagnetization field, the target skyrmion is no longer the ground state (Fig. S5). In order to quantify the effect of the demagnetization field, we used different values of saturation magnetization to tune the demagnetization energy, while keeping other interactions unchanged. As before, the system was relaxed from different initial states and the energies of the final states were compared. Fig S8 shows that the helical state has lower energy than the target skyrmion when the demagnetization field is 0.4 times as small as the original value. The variation in spin configuration along the z direction is identified directly in the simulation [27], as shown in Fig 4c. A slight twist of the magnetization at the top and bottom surfaces occurs because the presence of such a twist normal to the surface saves energy from the DM interaction along the z direction. In comparison, the lowest energy state for a 2D nanodisk is a complex helical state relaxed from a random initial magnetic configuration (Fig. S5).

Our numerical results confirm that two factors stabilize a zero-field target skyrmion, whose size is essentially determined by DM and FM couplings. Since the material parameters that are used here can be reduced to a small number of dimensionless variables, a scalar behavior is followed [31]. Our simulated results are then of universal significance for other helimagnetic materials. Moreover, skyrmion sizes in bulk are widely tunable from hundreds to tens of nm [32] and the temperature region of the skyrmion phase can be extended to room temperature [33]. The formation of a zero-field target skyrmion of much smaller size at room temperature is therefore anticipated.

The zero-field target skyrmion that we observe in nanodisks here is similar to the well-studied vortex in micron-sized elements of soft magnets [24]. Magnetic vortices have been investigated extensively, in the hope of building spintronic devices [34], because of the formation of four-fold-degenerate ground states with two polarities and two circularities. The present zero-field target skyrmion with a two-fold degeneracy can be an alternative to magnetic vortices in similar applications.

**Conclusions**

We have directly observed zero-field target skyrmions in a chiral magnetic material in a strongly confined nanodisk geometry using state-of-art off-axis electron holography. In the presence of an external magnetic field applied perpendicular to the nanodisk, the two types of target skyrmion can be switched. Our results are well reproduced and additional details are captured in numerical calculations. Our demonstration of the stability and switching of target skyrmions provides a solid physical basis for future skyrmion-based applications. The emergence of target skyrmions in other systems and their switching using different stimuli will be discussed in future works.

**Acknowledgments**

This work was supported by the Natural Science Foundation of China, Grant No. 51622105, 11474290; the Key Research Program of Frontier Sciences, CAS, Grant No. QYZDB-SSW-SLH009; the Youth Innovation Promotion Association CAS No. 2015267; the European Union through the Marie Curie Initial Training Network SIMDALEE2 and the European Research Council Seventh Framework Programme (FP7/2007-2013)/ ERC grant agreement number 320832. Numerical simulations at UNH were supported by the U.S. Department of Energy (DOE), Office of Science, Basic Energy Sciences (BES) under Award No. DE-SC0016424.

**References**

[1]  C. Binns, editor, *Nanomagnetism: Fundamentals and Applications* (Elsevier, Amsterdam Bonston Heidelberg London New York, n.d.).
[2]  S. Mühlbauer, B. Binz, F. Jonietz, C. Pfleiderer, A. Rosch, A. Neubauer, R. Georgii, and P. Böni, Science **323**, 915 (2009).
[3]  X. Z. Yu, Y. Onose, N. Kanazawa, J. H. Park, J. H. Han, Y. Matsui, N. Nagaosa, and Y. Tokura, Nature **465**, 901 (2010).
[4]  A. Fert, V. Cros, and J. Sampaio, Nat. Nanotechnol. **8**, 152 (2013).
[5]  N. Nagaosa and Y. Tokura, Nat. Nanotechnol. **8**, 899 (2013).
[6]  R. Ozawa, S. Hayami, and Y. Motome, Phys. Rev. Lett. **118**, 147205 (2017).
[7]  J. C. Gallagher, K. Y. Meng, J. T. Brangham, H. L. Wang, B. D. Esser, D. W. McComb, and F. Y. Yang, Phys. Rev. Lett. **118**, 027201 (2017).
[8]  U. K. Rößler, A. A. Leonov, and A. N. Bogdanov, J. Phys. Conf. Ser. **303**, 012105 (2011).
[9]  T. Shinjo, T. Okuno, R. Hassdorf, K. Shigeto, and T. Ono, Science **289**, 930 (2000).

[10] H. Du, W. Ning, M. Tian, and Y. Zhang, EPL Europhys. Lett. **101**, 37001 (2013).
[11] A. O. Leonov, U. K. Rößler, and M. Mostovoy, EPJ Web Conf. **75**, (2014).
[12] M. Beg, R. Carey, W. Wang, D. Cortés-Ortuño, M. Vousden, M.-A. Bisotti, M. Albert, D. Chernyshenko, O. Hovorka, R. L. Stamps, and H. Fangohr, Sci. Rep. 5, 17137 (2015).
[13] M. Beg, M. Albert, M.-A. Bisotti, D. Cortés-Ortuño, W. Wang, R. Carey, M. Vousden, O. Hovorka, C. Ciccarelli, C. S. Spencer, C. H. Marrows, and H. Fangohr, Phys. Rev. B **95**, 014433 (2017).
[14] M. Charilaou and J. F. Löffler, Phys. Rev. B **95**, 024409 (2017).
[15] S. Rohart and A. Thiaville, Phys. Rev. B **88**, 184422 (2013).
[16] Y. Liu, H. Du, M. Jia, and A. Du, Phys. Rev. B **91**, (2015).
[17] X. Zhao, C. Jin, C. Wang, H. Du, J. Zang, M. Tian, R. Che, and Y. Zhang, Proc. Natl. Acad. Sci. **113**, 4918 (2016).
[18] X. Yu, J. P. DeGrave, Y. Hara, T. Hara, S. Jin, and Y. Tokura, Nano Lett. **13**, 3755 (2013).
[19] H. Du, R. Che, L. Kong, X. Zhao, C. Jin, C. Wang, J. Yang, W. Ning, R. Li, C. Jin, X. Chen, J. Zang, Y. Zhang, and M. Tian, Nat. Commun. **6**, 8504 (2015).
[20] C. Jin, Z.-A. Li, A. Kovács, J. Caron, F. Zheng, F. N. Rybakov, N. S. Kiselev, H. Du, S. Blügel, M. Tian, Y. Zhang, M. Farle, and R. E. Dunin-Borkowski, Nat. Commun. in press, (2017).
[21] P. A. Midgley and R. E. Dunin-Borkowski, Nat. Mater. **8**, 271 (2009).
[22] A. Bogdanov and A. Hubert, J. Magn. Magn. Mater. **195**, 182 (1999).
[23] A. B. Butenko, A. A. Leonov, A. N. Bogdanov, and U. K. Rößler, Phys. Rev. B **80**, 134410 (2009).
[24] M. Kammerer, M. Weigand, M. Curcic, M. Noske, M. Sproll, A. Vansteenkiste, B. Van Waeyenberge, H. Stoll, G. Woltersdorf, C. H. Back, and G. Schuetz, Nat. Commun. **2**, 279 (2011).
[25] P. Milde, D. Köhler, J. Seidel, L. M. Eng, A. Bauer, A. Chacon, J. Kindervater, S. Mühlbauer, C. Pfleiderer, S. Buhrandt, C. Schütte, and A. Rosch, Science **340**, 1076 (2013).
[26] X. Z. Yu, N. Kanazawa, Y. Onose, K. Kimoto, W. Z. Zhang, S. Ishiwata, Y. Matsui, and Y. Tokura, Nat. Mater. **10**, 106 (2011).
[27] F. N. Rybakov, A. B. Borisov, and A. N. Bogdanov, Phys. Rev. B **87**, 094424 (2013).
[28] F. N. Rybakov, A. B. Borisov, S. Blügel, and N. S. Kiselev, Phys. Rev. Lett. **115**, (2015).
[29] A. O. Leonov, Y. Togawa, T. L. Monchesky, A. N. Bogdanov, J. Kishine, Y. Kousaka, M. Miyagawa, T. Koyama, J. Akimitsu, T. Koyama, K. Harada, S. Mori, D. McGrouther, R. Lamb, M. Krajnak, S. McVitie, R. L. Stamps, and K. Inoue, Phys. Rev. Lett. **117**, (2016).
[30] A. O. Leonov, J. C. Loudon, and A. N. Bogdanov, Appl. Phys. Lett. **109**, 172404 (2016).
[31] H. Du, W. Ning, M. Tian, and Y. Zhang, Phys. Rev. B **87**, 014401 (2013).
[32] N. Kanazawa, S. Seki, and Y. Tokura, Adv. Mater. 1603227 (2017).
[33] Y. Tokunaga, X. Z. Yu, J. S. White, H. M. Rønnow, D. Morikawa, Y. Taguchi, and Y. Tokura, Nat. Commun. **6**, 7638 (2015).
[34] S. Wintz, V. Tiberkevich, M. Weigand, J. Raabe, J. Lindner, A. Erbe, A. Slavin, and J. Fassbender, Nat. Nanotechnol. **11**, 948 (2016).